\documentclass[twocolumn,prd,preprintnumbers,amsmath,amssymb,floatfix,nofootinbib]{revtex4}

\usepackage{graphicx}
\usepackage{dcolumn}
\usepackage{bm}
\usepackage{verbatim}
\usepackage{tabularx}
\usepackage{color}

\preprint{MCTP-17-02}
\preprint{DESY-17-031}

\newcommand{\nc}{\newcommand}

\nc{\beq}{\begin{equation}} \nc{\eeq}{\end{equation}}
\nc{\beqa}{\begin{eqnarray}} \nc{\eeqa}{\end{eqnarray}}
\nc{\bea}{\begin{eqnarray}} \nc{\eea}{\end{eqnarray}}
\nc{\barray}{\begin{eqnarray}} \nc{\earray}{\end{eqnarray}}
\nc{\barrayn}{\begin{eqnarray*}} \nc{\earrayn}{\end{eqnarray*}}
\nc{\ra}{\rightarrow}
\nc{\lsim}{\begin{array}{c}\,\sim\vspace{-21pt}\\< \end{array}}
\nc{\gsim}{\begin{array}{c}\sim\vspace{-21pt}\\> \end{array}}
\nc{\Tr}{{\rm Tr}} \nc{\slsh}{\slash\hspace*{-0.22cm}}

\def\be{\begin{equation}}
\def\ee{\end{equation}}
\def\bea{\begin{eqnarray}}
\def\eea{\end{eqnarray}}
\def\bit{\begin{itemize}}
\def\eit{\end{itemize}}

\def\xfb{\, {\rm fb}}

\newcommand{\gev}{{\rm GeV}}
\newcommand{\tev}{{\rm TeV}}

\nc{\infinity}{\infty} \nc{\mc}{\mathcal} \nc{\M}{\mathcal{M}}

\def\to{\rightarrow}

\begin{document}

\title{Establishing the Isolated Standard Model}

\author{James D. Wells$^{a,b}$, Zhengkang~Zhang$^{a,b}$, Yue~Zhao$^{a}$}

\affiliation{
${}^{(a)}$Michigan Center for Theoretical Physics (MCTP), University of Michigan, Ann Arbor, Michigan 48109, USA
\\
${}^{(b)}$Deutsches Elektronen-Synchrotron (DESY), 22607 Hamburg, Germany
}

\date{\today}

\begin{abstract}
The goal of this article is to initiate a discussion on what it takes to claim ``there is no new physics at the weak scale,'' namely that the Standard Model (SM) is ``isolated.'' The lack of discovery of beyond the SM (BSM) physics suggests that this may be the case. But to truly establish this statement requires proving all ``connected'' BSM theories are false, which presents a significant challenge. We propose a general approach to quantitatively assess the current status and future prospects of establishing the isolated SM (ISM), which we give a reasonable definition of. We consider broad elements of BSM theories, and show many examples where current experimental results are not sufficient to verify the ISM. In some cases, there is a clear roadmap for the future experimental program, which we outline, while in other cases, further efforts -- both theoretical and experimental -- are needed in order to robustly claim the establishment of the ISM in the absence of new physics discoveries.
\end{abstract}

\maketitle


\section{Introduction}

Recent discovery of the Higgs boson without other new physics discoveries hints that Nature may be described by the Isolated Standard Model (ISM), where additional exotic states, if they exist, that couple to Standard Model (SM) particles are not in the neighborhood of the weak scale. Many beyond the Standard Model (BSM) theories that are invoked to solve outstanding problems in physics are implicitly assumed to be incompatible with the ISM. These include explanations of dark matter, new states that aid gauge coupling unification,  explanations of the baryon asymmetry, models to account for flavor hierarchies, and, most directly, solutions to the naturalness/fine-tuning problem (see, e.g., \cite{Csaki:2016kln,Allanach:2016yth}). If the ISM could be robustly established, it would have a profound impact on how we go about addressing these problems in the future.

Rather than making arbitrary claims about whether we have established, or will be able to establish the ISM, we would like to adopt a well-defined quantitative approach. Let us quantify our intuition about ``isolation'' in the following way. We define the SM to be isolated if there is no heavy new particle within an order of magnitude of the weak scale that couples to any SM particle with ${\cal O}(1)$ coupling, nor light new particle that couples to a SM particle with ${\cal O}(0.1)$ coupling. Also, the stronger the coupling to non-SM states, the less isolated the SM should be. Therefore, we write
\beq
d_i \equiv \frac{\max\{1,M_i/174~\text{GeV}\}}{\eta_i} \,,
\label{ISMpre}
\eeq
as a distance measure from the SM-only Lagrangian to a BSM interaction vertex $i$. The latter involves one or more non-SM particles, the heaviest of which has mass $M_i$, and a dimensionless coupling with magnitude $\eta_i$ (in units of $\hbar^{-1/2}$). We have used $m_t^\text{pole} \simeq 174~\gev$ as a representative of the weak scale. If an interaction vertex $i$ involves a dimensionful coupling, we normalize it by powers of 174~GeV. We say that interaction vertex $i$ (or coupling $\eta_i$) is isolated from the SM if $d_i>10$, and the full SM is isolated if
\beq
\min_{{\rm all}\, i} \{d_i\} > 10 \quad \Leftrightarrow \quad \text{The SM is isolated}.
\label{ISM}
\eeq
%

\begin{figure}[t]
\begin{center}
\includegraphics[width=0.48\textwidth]{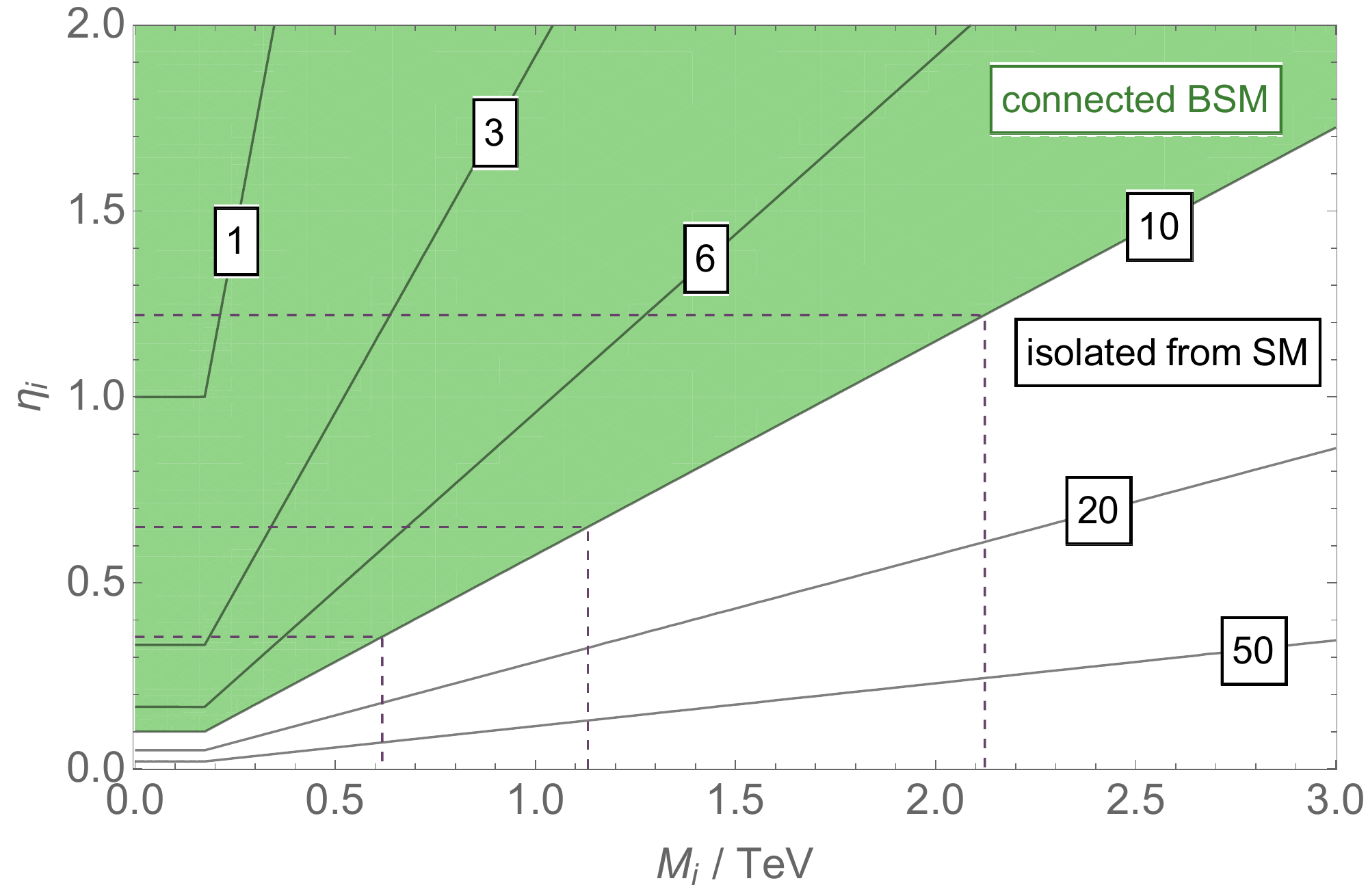}
\hspace*{0.35cm} \caption{Contours of $d_i$, a distance measure from the SM Lagrangian to a BSM interaction vertex $i$ as defined in Eq.~\eqref{ISMpre}, in the mass-coupling plane. To quantify our intuition about ``isolation'' of the SM, we define ``connected'' (i.e.\ non-isolated) BSM theories to be those with at least one interaction vertex with $d_i<10$, i.e.\ in the green shaded region. Dashed lines indicate the required mass exclusion limits of new particles in order to establish isolation from a BSM coupling of size $g_3, g_2, g_1$, respectively, the SM gauge couplings at $\mu=m_Z$.
} \label{fig:dicontour}
\end{center}
\end{figure}

Of course, our approach to quantifying ``isolation'' is by no means the only reasonable one. The spirit is to set up a simple criterion that agrees with intuition, as is demonstrated by the $d_i$ contours in the mass-coupling plane in Fig.~\ref{fig:dicontour}. For convenience we define a {\it connected BSM} theory to be any theory with at least one low-mass, well-coupled exotic state that renders the SM not isolated. Examples of connected BSM theories include supersymmetry with weak-scale superpartners, a light $Z'$ coupled to SM fermions, a scalar singlet that mixes sufficiently strongly with the SM Higgs boson, and an infinite number of other theories. By this definition all connected BSM theories are within the {\it isolation radius}, and all other BSM theories with all their states too massive or too weakly coupled are outside the {\it isolation radius}. Of course, if Nature is only the SM and nothing else, that also classifies under the ISM.

To prove that the SM is isolated requires showing that all possible connected BSM theories are false. This is a daunting task, and perhaps impossible, but it must be emphasized that this is often what is implied by the claim ``there is no new weak-scale physics." Showing all connected BSM theories are false is equivalent to showing how all connected BSM theories are visible to our experiments in at least one way that violates SM expectations. Therefore, making strides toward establishing the ISM requires analysis of a wide variety of connected BSM theories to determine if they are visible to current experiment and if not, what future experiment is needed to make them visible.

There are at least two important challenges to this very extensive goal. First, it is impossible to be absolutely comprehensive on the BSM theories to be analyzed. To make progress, we shall consider broad elements of BSM theories --- couplings to SM gauge bosons, to SM fermions, and to the SM Higgs boson --- and ask how they could be discerned by experiment. To some extent, this reflects how we usually, and perhaps implicitly, think about BSM physics.

A second challenge is that the interpretation of null experimental results as exclusion limits can be subtle and rely on assumptions. Take hadron colliders as an example. In many scenarios the existence of a connected BSM particle can be hidden if it decays into final states that have large backgrounds. One extreme example is when the new physics particle has a large exotic decay branching ratio into several jets within the scale of the detector. Thus one cannot always fully rely on direct production to look for connected BSM theories.

On the other hand, if new physics particles exist and couple to the SM sector, but are hidden in large collider backgrounds, they may show themselves indirectly, such as through distortions of differential distributions of SM scattering processes, and modifications of the SM Higgs boson branching ratios. Such indirect probes have the advantage of being more model independent, since one does not rely on how the new physics particles decay in their corresponding direct production searches. It should be emphasized that when trying to establish the ISM,  the goal is to find methods of discerning non-SM signals {\it of any kind} for the vast array of connected BSM theories. The goal is not necessarily to know exactly what is causing the deviation from the SM. In this sense, indirect probes are just as acceptable as direct probes when confronting
connected BSM theories to data in order to establish the ISM.

In what follows, we discuss several examples of BSM theories featuring each of the broad elements mentioned above, and analyze the status of establishing isolation in these specific contexts. As we will see, the LHC has made much progress, but we are still very far from establishing the ISM. We also discuss some of the existing ideas on how to make additional progress by future collider experimental probes, which may or may not be sufficient. Throughout the discussion the reader should keep in mind that the goal is to establish the ISM, which means that the most challenging signatures are the most important ones to point out. We end with a summary of our conclusions.

\section{Couplings to SM Gauge Bosons}

New particles coupling to SM gauge bosons are ubiquitous in BSM theories. They often come in the form of heavy partners of SM particles, as the latter are embedded in a deeper theoretical structure at higher energy scales. By naturalness or other arguments, it is quite conceivable that at least some of them are not too far isolated from the electroweak scale, motivating searches at high-energy hadron and lepton colliders.

Assuming minimal coupling, the BSM interaction vertex of interest here  involves new colored particles and the SM gluon, or new electroweak-charged particles and the SM electroweak gauge bosons. $\eta_i$ in Eq.~\eqref{ISMpre} is the corresponding SM gauge coupling (additionally multiplied by the hypercharge in the case of $U(1)_Y$-charged particles)\footnote{Note that for the non-abelian groups $SU(3)_c$ and $SU(2)_L$, we define $\eta_i$ in Eq.~\eqref{ISMpre} to be the gauge coupling independently of specific representation.}, which we take to be renormalized at the weak scale. By our isolation criterion, the SM is isolated when there are no new colored particles with $M\lesssim2.1$~TeV, no new $SU(2)_L$-charged particles with $M\lesssim1.1$~TeV, and no new $U(1)_Y$-charged particles with $M\lesssim|Y| \cdot 600$~GeV; see Fig.~\ref{fig:dicontour}. Below we discuss in turn colored and electroweak-charged new particles. We give some examples of direct search strategies, as well as more model-independent indirect probes.

\subsection{New colored particles}

High-energy hadron colliders offer the best opportunities of uncovering BSM couplings to gluons, or establishing the isolation of such couplings from the SM. As a typical example, the scalar top partner, stop ($\tilde t$), in supersymmetry can be searched for via pair production at the LHC. The reach in $m_{\tilde t}$ depends on the SUSY spectrum, and results are usually presented in $m_{\tilde t}$-$m_\chi$ plane, with $\chi$ being the lightest neutralino. Among various search channels~\cite{StopHadATLAS,StopSemiATLAS,StopLepATLAS,StopHadCMS,StopSemiCMS,StopLepCMS}, the best constraint at present is $m_{\tilde t}\gtrsim850$~GeV when $m_\chi\lesssim200$~GeV, which is still far from the 2.1~TeV goal. High-luminosity LHC can push the lower bound on $m_{\tilde t}$ beyond 1~TeV assuming light neutralino~\cite{HL-ATLAS,HL-CMS}, while to establish isolation would require a higher-energy collider. For example, it is found in~\cite{Stop100} that a 100~TeV $pp$ collider with 3~ab$^{-1}$ integrated luminosity will exclude $\tilde t$ up to 8~TeV, as long as $m_{\tilde t}-m_\chi$ is not very close to the top quark mass.

Another example of new colored particle is a vector-like top partner ($T$), which generically appears in e.g.\ composite Higgs models. Depending on its charge under $SU(2)_L$, it may have sizable decay branching ratios to $Zt$, $Ht$ or $Wb$. With 13 fb$^{-1}$ data at 13 TeV LHC, several search channels~\cite{VLTATLAS1,VLTATLAS2} give comparable limits and excludes $m_T$ up to around 800 GeV. A high-energy run of the LHC at 33~TeV would allow $T$ to be discovered up to about 2.5~TeV with 3~ab$^{-1}$ data, and with lepton+jet and multi-lepton channels combined~\cite{VLT33}.

We see that while present data cannot yet fully explore new colored particles such as $\tilde t$ and $T$ within the isolation radius of 2.1~TeV, future experiments show great prospects by continuing direct searches for such particles. Nevertheless, it should be kept in mind that no conclusions can be drawn from direct searches without making particular assumptions on the mass spectrum and/or decay channels. Simple model-building variations (see e.g.~\cite{Strassler:2006qa,StealthSUSY,Dobrescu:2016pda}) can easily render the signals invisible, and make it perhaps impossible to fully establish the isolation of BSM couplings to gluons via direct searches.

In this regard, it is worth considering indirect probes which may offer complementary information. For example, running of the strong coupling $g_3$ is sensitive to the presence of new colored particles above their mass thresholds, independent of their decay patterns~\cite{gluino51,R32}. Here, BSM contributions to the beta function are characterized by an effective multiplicity $n_3^\text{eff}$ under $SU(3)_c$, defined by
\beq
\Delta\beta_3 = \frac{d}{d\ln\mu}\left[g_3(\mu) -
g_3^\text{SM}(\mu)\right] \equiv \frac{g_3^3}{24\pi^2} n_3^\text{eff} \,,
\label{eq:n3eff}
\eeq
for $\mu>M$, the mass of the heavy colored particle(s). With this definition, we have, e.g.
\beqa
&& \text{(Majorana) gluino:} \quad n_3^\text{eff} = 3 \,, \\
&& \text{One degenerate squark generation:} \quad n_3^\text{eff} =
1 \,.\quad
\eeqa

In this framework, Ref.~\cite{R32} shows that, when the CMS 7~TeV analysis of inclusive 3-to-2-jet cross section ratios~\cite{CMSR32} is interpreted as a measurement of running strong coupling up to $\sim900$~GeV, bounds can be set in the $(M, n_3^\text{eff})$ plane. It is found that (degenerate) new physics with $n_3^\text{eff}=3(6)$, which is the case of a gluino (the full MSSM), is excluded up to 280(450)~GeV at 95\%~C.L. While these numbers are still far from the isolation radius of 2.1~TeV, it is hopeful that higher-energy hadron colliders, especially a future 100~TeV $pp$ collider, can make significant progress toward this target. Given the possibility that connected BSM theories featuring new colored particles may well be invisible to direct searches, a dedicated study of the potential reach of future colliders via such powerful indirect probes is highly warranted.

\subsection{New electroweak-charged particles}

\begin{table*}
\onecolumngrid

\begin{center}
\begin{tabular}{|c|c|c|c|}
\hline
 & Mono-jet & Disappearing track & Drell-Yan (running $g_2,g_1$) \\
 & HL-LHC (100~TeV)~\cite{EWdirect} & HL-LHC (100~TeV)~\cite{EWdirect} & HL-LHC\,+\,100~TeV~\cite{AGRW} \\
\hline
Pure wino & 0.3~(1.4)~TeV & 0.5~(3.5)~TeV & 1.3~TeV \\
\hline
Pure Higgsino & 0.2~(0.9)~TeV & 0.2~(0.8)~TeV & invisible \\
\hline
\end{tabular}
\caption{\label{tab:WinoHiggsino}
Projected 2$\sigma$ exclusion limits for pure wino and Higgsino, from different search strategies at hadron colliders.}
\end{center}
\twocolumngrid
\end{table*}

Compared with colored particles discussed above, purely electroweak-charged states are generically more difficult to search for at hadron colliders. We consider as examples (almost) pure wino and Higgsino, which can  be the only BSM states in the vicinity of the electroweak scale, as predicted in many models including (mini-)split SUSY~\cite{Wells:2003tf,ArkaniHamed:2004fb,ArkaniHamed:2004yi,Wells:2004di,Giudice:2004tc,Arvanitaki:2012ps,ArkaniHamed:2012gw}. They are especially challenging to look for because the mass splitting between charged and neutral states is usually very small. Thus the decay products from these almost degenerate states can be too soft to be triggered. In this case, one has to rely on a hard initial state radiation (ISR) for triggering. Thus a mono-jet search can be used to probe this degenerate spectrum. On the other hand, due to the small splitting, the decay of charged state can be very displaced in some parts of the parameter space, allowing for disappearing track searches. These two search strategies have been carefully studied in~\cite{EWdirect}, in the contexts of the HL-LHC (14~TeV with 3000~fb$^{-1}$ integrated luminosity) and a future 100~TeV $pp$ collider. In particular, $1\%$ systematic uncertainty has been assumed for the mono-jet channel, while an optimistic estimation of the dominant background from mis-measured low-$p_T$ tracks has been adopted. We quote their results in the second and third columns of Table~\ref{tab:WinoHiggsino}. We see that HL-LHC is not good enough to establish isolation but $100\,\tev$ collider will greatly extend the mass reach.

A third, indirect, strategy, studied in detail in~\cite{AGRW} (see also~\cite{RainwaterTait07}), is to detect effects of new electroweak-charged particles on the running of electroweak couplings $g_2, g_1$. Analogous to Eq.~\eqref{eq:n3eff} for the case of couplings to gluon, one can define the effective multiplicities $n_2^\text{eff}$, $n_1^\text{eff}$ under $SU(2)_L$ and $U(1)_Y$, respectively, and set bounds in the $(M, n_2^\text{eff})$ and $(M, n_1^\text{eff})$ planes. This is done in~\cite{AGRW}, where invariant mass distribution in the neutral Drell-Yan process $pp\to Z^*/\gamma^*\to l^+l^-$ and transverse mass distribution in the charged Drell-Yan process $pp\to W^*\to l\nu$ are interpreted as running electroweak coupling measurements. It is found that current capabilities of the LHC are far from establishing isolation of electroweak states using this technique. However, future experiments may have a significant impact. In particular, with HL-LHC and 100~TeV combined, $2\sigma$ exclusion of $SU(2)_L$-charged ($U(1)_Y$-charged) new particles below $\sim1.3$~TeV ($\sim700$~GeV) is possible if $n_2^\text{eff}\gtrsim2$ ($n_1^\text{eff}\gtrsim8$). The sensitivity to $n_2^\text{eff}, n_1^\text{eff}$ is bounded below those masses, and is lost rapidly above them. For our examples, we have
\beqa
&& \text{Wino (Majorana):}\quad n_2^\text{eff}=2\,, \quad n_1^\text{eff} =0\,, \\
&& \text{Higgsino (Dirac):}\quad n_2^\text{eff}=\frac{1}{2}\,, \quad
n_1^\text{eff} =\frac{1}{4}\,, 
\eeqa 
leading to the last column of Table~\ref{tab:WinoHiggsino}.

From Table~\ref{tab:WinoHiggsino} we see that all three search strategies show promising prospects of establishing the isolation of gauge interactions of a pure wino, but only if we have inputs from a 100~TeV $pp$ collider. On the other hand, a pure Higgsino within the isolation radius of $1.1$~TeV may still escape detection, even with the most powerful hadron machine ever seriously considered. Such a low-effective-multiplicity state will however be easily visible at an $e^+e^-$ collider, once sufficient energy is reached. In fact, for the purpose of just seeing the BSM effects, the center-of-mass energy requirement is somewhat lower than the pair production threshold. This is because virtual effects of new physics coupling to electroweak gauge bosons can modify $e^+e^-\to f\bar f$ differential cross sections. For example, it is found in~\cite{JapFLC} that, under optimistic assumptions on uncertainties, to exclude a 1.1~TeV Higgsino, one only needs $\sqrt{s}\gtrsim2(1.2)$~TeV, assuming 1(10)~ab$^{-1}$ integrated luminosity. We see that from the point of view of establishing the ISM, a high-energy $e^+e^-$ collider will provide crucial information on BSM couplings to electroweak gauge bosons, complementary to that from hadron colliders.

\section{Couplings to SM Fermions}

We next consider the case of new bosonic fields mediating interactions among SM fermions. Their presence is predicted by many well-motivated BSM theories such as composite Higgs and extra dimensions. According to our criterion, when the bosonic mediator has a mass $M>174~\gev$, we shall say its interactions with SM fermions are isolated if 
\be 
\frac{d_i}{10} = \frac{1}{|\eta_i|}\cdot \frac{M}{1.74~\text{TeV}} >1 ~~\text{(isolated)} \,, 
\label{dferm} 
\ee 
for all vertices $i$. We will investigate in the following how far we are from establishing Eq.~\eqref{dferm} in the contexts of three well-motivated examples --- $Z'$, axigluon (a.k.a.~$G'$) and leptoquarks (see e.g.~\cite{ZprimeReview,Frampton:1987dn,LQReview}) --- and if future experiments can bring us closer to that goal with several existing (direct and indirect) search strategies.

\subsection{$Z'$ and $G'$}

Resonance searches for $Z'$ and $G'$ particles can be extremely powerful in establishing Eq.~\eqref{dferm}. For example, a $Z'$ coupling to fermions identically as the SM $Z$ boson (for which all $|\eta_i|<1$) has already been excluded by LHC dilepton resonance searches~\cite{DilepResATLAS,DilepResCMS} up to $3\sim4$~TeV. The more challenging scenario is that of a leptophobic $Z'$, or a $G'$. But even in those cases, dijet resonance searches~\cite{DijetResATLAS,DijetResCMS} can set exclusion limits of at least 2.5~TeV. We therefore conclude that current LHC data can already establish the isolation of $Z'$ and $G'$ couplings to SM fermions, {\it assuming} these new states are visible as narrow resonances in the dilepton or dijet channel. This latter assumption is, however, not generically satisfied in connected BSM theories. One can easily imagine scenarios where a $Z'$ or $G'$ has a large branching ratio into multi-jets, or a large total width due to invisible decay. To fully establish the isolation requires excluding  also such scenarios, which are very challenging, and likely inviable in direct resonance searches.

In light of this, it is equally important to consider off-resonance searches. For example, differential distribution measurements of dijet production at the LHC can reveal deviations from SM predictions, induced by an off-shell $Z'$ (leptophobic or not) or $G'$ mediating $q\bar q\to q\bar q$ scattering. An interesting limit to consider is $M\gg \sqrt{\hat s}$, where BSM-induced dijet events are more concentrated in the central region compared to SM backgrounds. This can be quantified by the differential distribution $\frac{d\sigma}{d\chi}$ in certain dijet invariant mass bins. Here $\chi\equiv e^{|y_1-y_2|}$, with $y_{1,2}$ being the rapidities of the two jets. Many studies in the literature adopt an effective field theory framework, where four-quark contact interaction (CI) dimension-six operators are considered --- they provide a leading-order approximation to the effects of heavy $Z'$ and $G'$ coupling to quarks, when $M\gg \sqrt{\hat s}$. Limits are often quoted as lower limits on the suppression scales of CI operators, which can be directly translated into our distance measure $d_i$. Specifically, to establish the isolation requires 
\beq 
\Lambda_{ij} = \frac{M}{\sqrt{|\eta_i\eta_j|}} = \frac{\sqrt{d_i d_j}}{10}\cdot 1.74~\text{TeV} > 1.74~\text{TeV} \,.\label{Lambdaij} 
\eeq 
Several experimental analyses in this framework have been done in the simplified scenario where certain sets of $\Lambda_{ij}$ are finite and identical~\cite{CICMS13,CIATLAS13}. A more detailed picture of constraints is obtained in~\cite{4q:flavored}, where it is found that bounds on individual $\Lambda_{ij}$'s from 7~TeV CMS data~\cite{CICMS7} range from 1.5 to 3~TeV (in our notation), as long as first-generation quarks are involved. By Eq.~\eqref{Lambdaij}, this is already very close to establishing the isolation.

Nevertheless, establishing the isolation of CIs is not equivalent to establishing the isolation of $Z'$ and $G'$ models, because CIs only provide a valid description of new physics in the regime $M\gg\sqrt{\hat s}$. In particular, the numbers quoted above are obtained by restricting to the highest dijet invariant mass bin $m_{jj}>3$~TeV~\cite{4q:flavored}. This is motivated by the fact that BSM effects as captured by CIs grow with $\sqrt{\hat s}$. But as a result, the bounds cannot be directly translated into $Z'$ and $G'$ models with lower masses. To address this issue, Ref.~\cite{4q:CIvsModels} makes use of ATLAS 7~TeV dijet data in a broader $m_{jj}$ window, from 1.2~TeV to 4~TeV~\cite{4q:ATLASdijet}, and quantifies differences between full and effective theory exclusions. The $F_\chi$ observable, defined as the fraction of events in the central region $1<\chi<3.32$, is used in the analysis. For the benchmark models considered, where a $Z'$ or $G'$ couples universally to SM quarks via the vector current with coupling $\eta$, a simple empirical relation is found in~\cite{4q:CIvsModels} between exclusion curves in the $(M,\eta)$ plane derived from the full theory vs.\ the CI framework:
\beqa
&& \text{CI:} \qquad \eta < \frac{M}{\Lambda_\text{min}} \,, \\
&& \text{Full:} \qquad \eta < \left(1+\frac{C^2}{M^2}\right) \frac{M}{\Lambda_\text{min}} \,,
\label{eq:Full-CI}
\eeqa
where $C$ is a collider energy-dependent constant that is conjectured to be approximately universal for different models. In particular, it is found to be $\sim$1.3(2.2)~TeV at the 7(14)~TeV LHC for both $Z'$ and $G'$ benchmark models considered.

\begin{figure}[t]
\begin{center}
\includegraphics[width=0.48\textwidth]{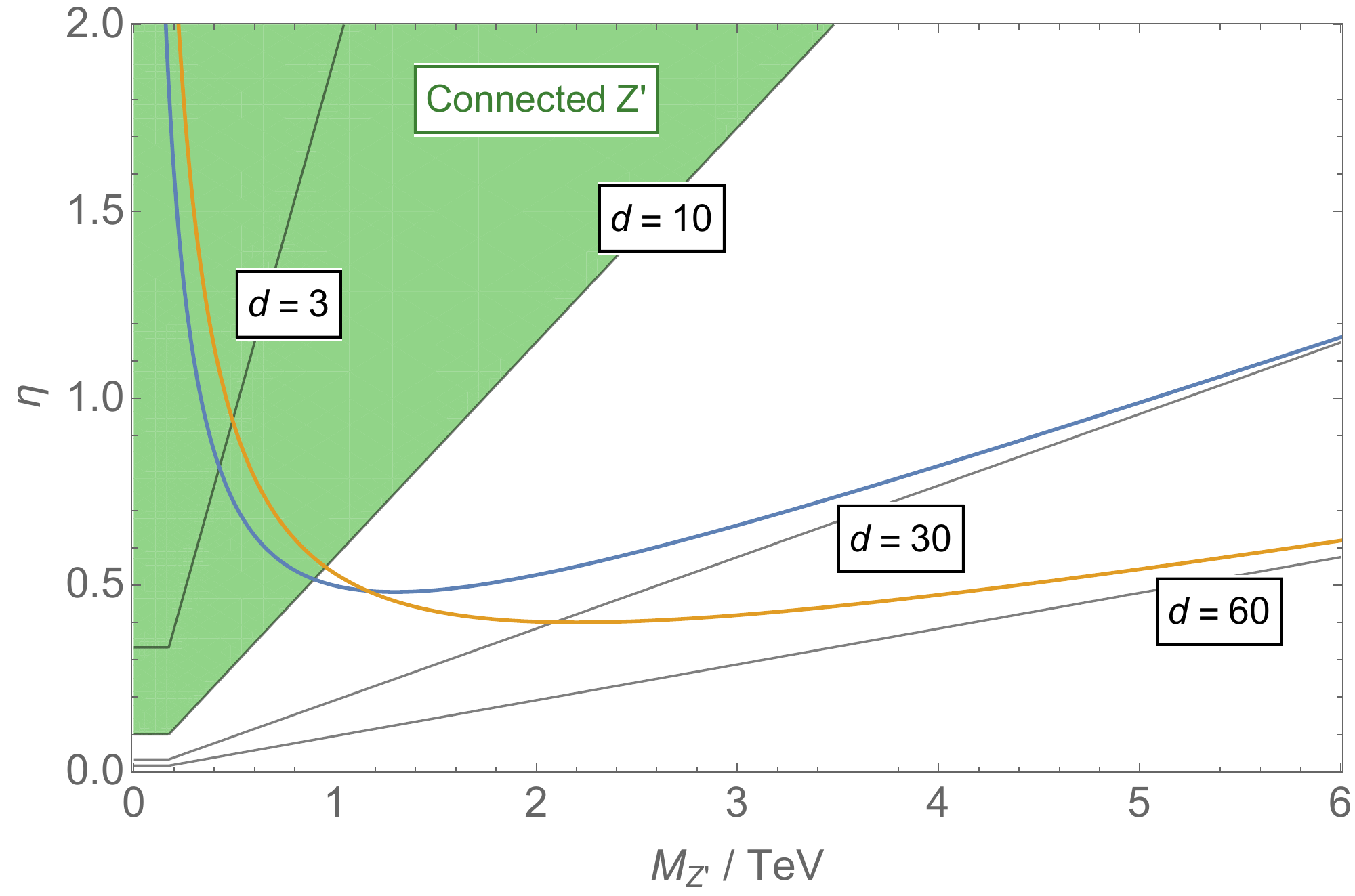}
\hspace*{0.35cm} 
\caption{
Existing (7~TeV, 4.8~fb$^{-1}$, blue) and projected (14~TeV, 100~fb$^{-1}$, orange) LHC exclusion limits on a benchmark $Z'$ model coupling universally to SM quarks via vector current from off-resonance search, in the $Z'$ mass-coupling plane, following the empirical parameterization of Eq.~\eqref{eq:Full-CI} found in~\cite{4q:CIvsModels}, with $\Lambda_\text{min}=5.4(11)$~TeV, $C=1.3(2.2)$~TeV for 7(14)~TeV. The parameter space region where the $Z'$ theory is connected to the SM (green shaded) is difficult to fully exclude with existing search strategy, and going to higher energy, despite doubling the reach in the distance measure in the high mass regime, does not close the sub-TeV blind spot.
} \label{fig:4q}
\end{center}
\end{figure}

Eq.~\eqref{eq:Full-CI} allows us to translate lower bounds on the suppression scale $\Lambda_\text{min}$ derived from a CI analysis to exclusion curves in the $(M,\eta)$ plane for $Z'$ and $G'$ models. We take the benchmark $Z'$ model in~\cite{4q:CIvsModels} for example, for which the 7~TeV LHC constraint reads $\Lambda_\text{min}\simeq5.4$~TeV in our notation\footnote{This is stronger than bounds obtained in~\cite{4q:flavored}, quoted previously, because more than one CI operators are generated in the model, and dijet data in more $m_{jj}$ bins are considered.}. We see from Fig.~\ref{fig:4q} that while a large portion of the parameter space where the $Z'$ theory is connected to the SM has already been excluded, a blind spot remains for $M\lesssim0.9$~TeV. Interestingly, while going from 7 to 14~TeV will double the reach in the CI suppression scale to $\Lambda_\text{min}\simeq11$~TeV, hence doubling the reach in the distance measure from $d\simeq30$ to $d\simeq60$ in the high mass regime, it does not help to close this sub-TeV blind spot. This is because the idea underlying current search strategies relies on new particles being heavy enough to induce dijet angular distributions distinguishable from SM backgrounds. Clearly, to fully establish the isolation of $Z'$, $G'$ couplings to SM quarks in a model-independent way (namely via off-resonance searches), it is imperative to go beyond strategies currently employed at the LHC.

\subsection{Leptoquarks}

Unlike $Z'$ and $G'$, leptoquarks cannot be produced as $s$-channel resonances at either hadron or lepton colliders. At the LHC, they are typically searched for via pair production. Such processes mainly rely on leptoquarks' gauge coupling to the SM gluon, and therefore do not suffice to fully establish the ISM,  which additionally requires leptoquarks' couplings to SM fermions to have $d_i>10$. For example, a scalar leptoquark, assumed to decay exclusively into a first or second generation charged lepton and a quark, has been excluded up to about 1.1~TeV~\cite{LQATLAS,LQCMS}. This cannot establish isolation of its couplings to SM fermions if the latter are $\gtrsim0.5$. Even with more data, it would be difficult to exclude connected BSM theories containing heavy leptoquarks with large couplings to SM fermions via pair production.

To gain direct access to leptoquarks' couplings to SM fermions, one can look at, e.g., single leptoquark production at the LHC~\cite{LQsingle}. However, a more powerful probe is offered by indirect searches via dilepton spectra in Drell-Yan processes, which can be modified by $t$-channel leptoquark exchange. For example, Ref.~\cite{LQRaj} considers four benchmark models, where a scalar leptoquark couples to first/second-generation lepton and right-handed up/down quark, and derives exclusion limits on the mass-coupling plane from 8~TeV LHC. In particular, the ATLAS measurement of dilepton invariant mass distribution~\cite{DYATLAS} and CMS measurement of forward-backward asymmetry~\cite{DYCMS} are used. It is found that, for a 1.05(2.5)~TeV leptoquark, upper bounds on its coupling to SM fermions in the benchmark models range from 0.42 to 0.72 (0.92 to 1.5). This is already very close to establishing isolation, namely Eq.~\eqref{dferm}, of these benchmark models. With the same observables, high-luminosity LHC will push the limits well beyond the isolation radius. It is therefore reasonable to expect that a full coverage of connected leptoquark interactions with SM fermions will be possible, even beyond the benchmark models considered in~\cite{LQRaj}.

\section{Couplings to SM Higgs Boson}

The Higgs sector has been one of the most studied fields in theoretical physics given the late confirmation even that a Higgs boson existed. There are numerous ideas that have sprung up over the years of how to break electroweak symmetry in a natural way. There is no way to study all possible theories of the Higgs sector that lead to a non-isolated SM. Nevertheless, we can discuss a few examples that give some guidance to how well experiment would be able to find all states coupling to the Higgs boson in a connected BSM theory, which would in turn obviate the ISM, or put it on firmer ground if no discovery is made.

\subsection{Extra Higgs doublet}

The first example is considering a Higgs sector with another Higgs doublet. There are many forms of the two Higgs doublet model, such as type I and type II and many other variants.  Suffice it to say that when there is another Higgs boson doublet in the spectrum, the total number of propagating degrees of freedom go from one in the SM (the Higgs boson) to five -- two neutral CP-even Higgs bosons ($H$ and $h$, with the latter identified with the $125\,\gev$ discovery), one CP-odd Higgs boson ($A$) and two charged Higgses ($H^\pm$). What is important to note here is that all of these extra Higgs bosons couple not only to the Higgs boson but to some of the SM states, including the gauge bosons and the fermions. In addition to producing single heavy neutral Higgs bosons through loop-induced gluon fusion at hadron colliders, an important well-established strategy to discover these Higgs bosons is through tree-level pair production via its gauge interactions and then decays via its Yukawa interactions with fermions.

The  limits on exotic Higgs bosons also depend on the flavor structure of their Yukawa couplings. One of the most non-intrusive ways to have a second Higgs doublet is for one doublet to couple to up-type fermions ($H_u$) and the other to down-type fermions ($H_d$). This is the ``type II" approach, and it is the standard construction in minimal supersymmetric theories. In that case, difficult flavor problems are rather easily avoided for exotic Higgs bosons that are well within the isolation definition. Limits are then primarily a function of the exotic Higgs boson mass and the value of $\tan\beta$, which is the ratio of vacuum expectation values of the two Higgs doublets:  $\tan\beta=\langle H_u\rangle/\langle H_d\rangle$.

Searching for the charged Higgs boson is one of the key probes of this extra-Higgs doublet scenario. Current limits at the LHC are not particularly strong with respect to our definition of SM isolation. For example, with $13.2\xfb^{-1}$ integrated luminosity at $\sqrt{s}=13\,\tev$, limits on the charged Higgs boson are $m_{H^+}>400\,\gev$ provided that $\tan\beta>1$ and $H^\pm$ decays primarily into top and bottom quarks~\cite{Atlas:2016-089}, which are reasonable assumptions in many detailed models. The difficulty of the search is illustrated well by the large region of parameter space that cannot yet be probed (see Fig.~13 of~\cite{Atlas:2016-089}).

Limits on the entire heavy Higgs sector from ATLAS and CMS searches, including searches for $A$ and $H$ in addition to $H^\pm$ are consistently within the few hundred GeV region. These search results are summarized well in Fig.~2 of~\cite{Ohman:2016}.

Future enhanced sensitivity to heavy Higgses can be accomplished at a higher energy $pp$ or $e^+e^-$ collider. For example, an $e^+e^-$ collider can find heavy Higgs bosons through $e^+e^-\to AH$ up to nearly the kinematic limit~\cite{Coniavitis:2007me}. A $3\,\tev$ CLIC with $2\, {\rm ab}^{-1}$ of integrated luminosity, for example, can make precision studies of charged Higgs and pseudo-scalar states at least up to masses of $900\,\gev$~\cite{Linssen:2012hp}, and discovery for masses  up to $\sim 1.4\,\tev$.  This would well establish the existence of such particles if they are connected to the SM (non-isolated).

\subsection{Real scalar singlet}

The case of the two Higgs doublet model is one where it is expected in time that evidence for non-isolated states would show up. It may require a next generation collider, like CLIC, but the path is conceivable. On the other hand, there are other ideas where the path is not so easily seen. Let us go to our second example with the goal of identifying a connected BSM theory that would be very difficult for current generation and next generation experiment to discover.

The example theory for this is to have an additional electroweak singlet scalar. In this theory, the singlet $\phi$ has mass $m_\phi$ and couples to the Higgs boson through a gauge invariant operator $\frac{\lambda}{2} \phi^2 |H|^2$. The $\phi$ state may or may not couple to other exotic states in a hidden sector, and therefore, for example, may or may not be dark matter. All we postulate with regard to its coupling to the SM is its coupling to the Higgs. Let us now discuss how such a state could be found.

If it is light enough, $m_\phi<m_h/2$ the Higgs boson could decay $h\to \phi\phi$ with the $\phi$ being stable and hence invisible in the detector, or decaying promptly in some way to additional SM states. This implies an exotic branching fraction well above most limits. For example, the current limits on the Higgs boson invisible branching fraction is 0.24~\cite{CMS:2016-016}.  If we interpret $h\to \phi\phi$ as an invisible decay mode of the Higgs, the limit on the interaction coupling is approximately $\lambda < 10^{-2}$, which is well below the isolation requirement. This result is to be expected, since the Higgs boson width is accidentally very small and therefore can be overwhelmed from even weakly coupled exotics~\cite{Wells:2009kq}. This is therefore a case of data already covering well any new physics connected to the SM in that sector and mass range.

On the other hand, if $m_\phi>m_h/2$ the situation is much more difficult. Producing $\phi$ directly through SM processes may require that they be produced in pairs if $\phi$ is charged under a discrete symmetry (e.g., odd under $Z_2$), which is often the case for extra scalar theories. Pair production of $\phi$ scalars would be extremely difficult at the LHC. However, more efficient probes -- both direct and indirect -- are possible if the discrete symmetry is not present, or if $\phi$ originates from a real scalar singlet $S$ that couples to the SM Higgs via $\frac{1}{2}\lambda_{HS}|H|^2 S^2$ and has a vacuum expectation value $v_s$. In this latter case, we can write
\beq
H = \left(
\begin{matrix}
0 \\
\frac{v+h'}{\sqrt{2}}
\end{matrix}
\right),
\quad
S = \frac{v_s+\phi'}{\sqrt{2}} \,,
\eeq
in unitary gauge, with the mass eigenstates obtained by
\beq
\left(
\begin{matrix}
h \\
\phi
\end{matrix}
\right) = \left(
\begin{matrix}
\cos\alpha & -\sin\alpha \\
\sin\alpha & \cos\alpha
\end{matrix}
\right) \left(
\begin{matrix}
h' \\
\phi'
\end{matrix}
\right) .
\eeq 
This scenario has been studied extensively for the LHC and future colliders as well; see e.g.~\cite{Bowen:2007ia,Robens:2015gla,SingletFuture,Robens:2016xkb}. In the simplest case where $S$ is assumed to be odd under a $Z_2$ symmetry, all scalar couplings can be written in terms of $m_\phi$, $v_s$ and $\lambda_{HS}$. Focussing on the cubic interactions $\frac{1}{2}(\lambda_{h^2\phi}h^2\phi +\lambda_{h\phi^2}h\phi^2)$, with $\lambda_{h^2\phi}, \lambda_{h\phi^2}$ having dimensions mass$\,\times\,$coupling, we find that for $m_\phi\gg m_h$, 
\beqa
\min\{d_{h^2\phi}, d_{h\phi^2}\} &=& \frac{m_\phi}{\max\{|\lambda_{h^2\phi}|, |\lambda_{h\phi^2}|\}} \nonumber\\
&\simeq& \frac{m_\phi}{\lambda_{HS}\cdot\max\{v,|v_s|/2\}} \,.
\label{dsingletcubic}
\eeqa
This approximate expression holds up to ${\cal O}(1)$ factors also when $m_\phi\sim v$.

We see that for $m_\phi \sim v_s \sim v$, establishing isolation of the cubic interactions would require excluding $\lambda_{HS}\gtrsim0.1$. This is likely an achievable task once the universal Higgs coupling shift
\beq
\sin^2\alpha \simeq \lambda_{HS}^2\left(\frac{v \cdot v_s/2}{m_\phi^2-m_h^2}\right)^2 ,
\eeq
can be constrained at percent level, perhaps at a future lepton collider. Furthermore, in this low mass regime, direct searches for $\phi$ in di-Higgs and diboson channels will offer a powerful complementary probe at the LHC and future hadronic colliders~\cite{SingletFuture}.

On the other hand, for $m_\phi\gtrsim1$~TeV and $v_s\sim v$, an ${\cal O}(1)$ or larger value for $\lambda_{HS}$ would correspond to a distance measure less than 10 (meaning the SM is not isolated), as seen from Eq.~\eqref{dsingletcubic}. Nevertheless, the resulting universal Higgs coupling shift could easily be at the sub-per-mil level, which is likely beyond reach of present and future experiments. Also, direct searches for the $\phi$ resonance are expected to be less powerful in such high mass regime~\cite{SingletFuture}. It is interesting to see how a simple BSM theory that is quite intuitively connected to the SM (with a 1~TeV new particle and ${\cal O}(1)$ coupling) can easily evade detection even with tremendous experimental efforts.

As a final remark, our definition of isolation is agnostic as to how large corrections may be of the Higgs boson couplings to other SM states, since isolation is defined through how exotic states interact with the SM, and is not defined directly through how SM states interact with each other. However, connected new states, such as superpartners or composite states, indirectly may affect the interactions of the Higgs boson with other SM states. In fact, even if such new states are not directly accessible at the LHC, they can lead to Higgs coupling modifications as large as 10\% for vector bosons, and even larger for fermions~\cite{HiggsCouplingsHowWell}. It is this connection that makes precision Higgs coupling analysis relevant for establishing the ISM or to find evidence for a connected BSM theory. At the present time, the Higgs sector is subject to many uncertainties given that our measurements of  Higgs couplings to SM states are only at the ${\cal O}(15\%)$ level~\cite{Khachatryan:2016vau}. Thus, there are many such connected BSM theories  that we currently have no sensitivity to (e.g., giving Higgs coupling deviations of a few percent), but whose effects could be seen in future experiments, such as ILC or CLIC, that substantially reduces those uncertainties~\cite{Baer:2013cma,Abramowicz:2016zbo}. Meanwhile, much effort is needed from the theory community to bring down theoretical uncertainties to match the expected experimental precision~\cite{PrecHiggsTH1,PrecHiggsLat,PrecHiggsTH2}.

\section{Conclusions}

Lack of evidence of new physics at the LHC invites us to consider the possibility that the Standard Model is isolated from new physics at the weak scale. We have defined through Eqs.~(\ref{ISMpre}) and~(\ref{ISM}) the Isolated Standard Model (ISM) as a theory that contains the SM and no other weak-scale particles with non-small couplings to SM states. As we have argued, we are currently far from establishing the ISM. Establishing the ISM requires ruling out all ``connected BSM theories", which have at least one state with interactions not isolated from the SM. This is a daunting task that will require systematic thinking about all possible classes of connected BSM theories and their experimental consequences, and it will require  a substantial experimental program progressing into the future to search for them.

Weak-scale supersymmetry has been a classic example over the years of a connected BSM theory that was expected to be found at the LHC. At the present time weak-scale supersymmetry inside the {\it isolation radius} has not been ruled out. Gluino/squark bounds are not yet at $2.1\,\tev$, and Higgsino/wino/slepton bounds are far from the necessary $1.1\,\tev$ isolation criterion level, for example. In addition, even heavy Higgs mass bounds are far from excluding all connected exotic pseudo-scalars or charged Higgs bosons. Thus we have significant effort still to go to rule out the low-scale supersymmetric class of connected BSM theories by perhaps any reasonable criterion, including ours.

One can consider other motivated connected BSM theories, such as warped extra dimensions, little Higgs, and composite Higgs theories, to map out the parameter space where there exists points within the {\it isolation radius} (i.e., connected new states) that are not yet probed by experiment. Showing how future high-luminosity and high-energy runs of the LHC can plug the visibility gaps in the connected BSM theory parameter space will be needed, keeping in mind that the most stubbornly collider-invisible connected BSM theories are the roadblocks to establishing the ISM.  Some connected BSM theories will likely then need next-generation experiments such as a 100 TeV hadron collider or multi-TeV CLIC to be found or ruled out.  Either way, concluding this search to establish the ISM, or alternatively finding a connected BSM theory, would have deep implications for fundamental science.

\begin{acknowledgments}
This work was supported in part by the U.S.\ Department of Energy under grant DE-SC0007859 and the Humboldt Foundation.
\end{acknowledgments}


\end{document}